\documentclass[a4paper,11pt]{article}

\usepackage{amssymb,amsbsy,amsmath,amsthm}

\newcommand{\Real}{\mathop{\mathbb R}\nolimits}           
\newcommand{\posReal}{\mathop{\mathbb R}^+\nolimits}      
\newcommand{\Nat}{\mathop{\mathbb N}\nolimits}            

\newcommand{\ie}{\emph{i.e.}}                             
\newcommand{\eg}{\emph{e.g.}}                             
\newcommand{\cf}{\emph{cf.}}                              
 
\newcommand{\supp}{\mathop{\mathrm{supp}}\nolimits}       
\newcommand{\diag}{\mathop{\mathrm{diag}}\nolimits}       
\newcommand{\Dom}{\mathop{\mathrm{Dom}}\nolimits}	  

\newcommand{\layer}{\mathop{\mathcal{L}}\nolimits}        
\newcommand{\TotK}{\mathop{\mathcal{K}}\nolimits}         
\newcommand{\TotM}{\mathop{\mathcal{M}}\nolimits}         

\newcommand{\si}{\mathop{L^1}\nolimits}                   
\newcommand{\sii}{\mathop{L^2}\nolimits}                  
\newcommand{\Sobi}{\mathop{W^{1,2}}\nolimits}             
\newcommand{\sobi}{\mathop{W_0^{1,2}}\nolimits}           
\newcommand{\Comp}{\mathop{C_0^\infty}\nolimits}          
\newcommand{\Smooth}{C}                                   
\newcommand{\tangent}{\textrm{T}}			  

\newcommand{\Ass}{\mbox{$\langle\Omega0\rangle$}}
\newcommand{\AssFlat}{\mbox{$\langle\Sigma0\rangle$}}
\newcommand{\AssWidth}{\mbox{$\langle\Omega1\rangle$}}
\newcommand{\AssK}{\mbox{$\langle\Sigma1\rangle$}}
\newcommand{\AssM}{\mbox{$\langle\Sigma2\rangle$}}
 
\newtheorem{claim}{Claim}[section]
\newtheorem{thm}[claim]{Theorem}                          
\newtheorem{lemma}[claim]{Lemma}                          
\newenvironment{PROOF}
  {\begin{proof}[\textsc{Proof:}]}{\end{proof}}		  

\theoremstyle{definition}
\newtheorem*{rem}{Remark}           			  
\newtheorem*{rems}{Remarks}				  

\theoremstyle{remark}
\newtheorem{Ex}{EXAMPLE}  		  		  
		
\begin{document}
\title{\textbf{Bound States in Curved Quantum Layers}
}
\author{P.~Duclos$^{a,b}$,
        P.~Exner$^{c,d}$,
        and D.~ Krej\v{c}i\v{r}\'{\i}k$^{a,b,c,e}$}
\date{}
\maketitle
\vspace{-5.5ex}
{\small\em
\begin{description}
\item[$a)$] Centre de Physique Th\'eorique, CNRS,
            13288 Marseille-Luminy
\vspace{-1.2ex}
\item[$b)$] PHYMAT, Universit\'e de Toulon et du Var,
            83957 La Garde, France
\vspace{-1.2ex}
\item[$c)$] Nuclear Physics Institute, Academy of Sciences,
            25068 \v{R}e\v{z} near Prague
\vspace{-1.2ex}
\item[$d)$] Doppler Institute, Czech Technical University,
            B\v{r}ehov{\'a}~7, 11519 Prague
\vspace{-1.2ex}
\item[$e)$] Faculty of Mathematics and Physics, Charles University,
            V~Ho\-le\-\v{s}o\-vi\v{c}\-k\'ach~2, 18000 Prague, Czech Republic
\vspace{-1.2ex}
\end{description}
}
\begin{center}
duclos@univ-tln.fr, exner@ujf.cas.cz, krejcirik@ujf.cas.cz
\end{center}
\begin{abstract}
\noindent
We consider a nonrelativistic quantum particle constrained 
to a curved layer of constant width built over a non-compact surface 
embedded in~$\Real^3$. We suppose that the latter is endowed with 
the geodesic polar coordinates and that the layer has the hard-wall
boundary. Under the assumption that the surface curvatures vanish
at infinity we find sufficient conditions which guarantee the existence
of geometrically induced bound states. 

\medskip\noindent
\textbf{Key-Words:} waveguides, layers, constrained systems, 
Dirichlet La\-placian, bound states, surface geometry, curvature, 
integral curvatures, geodesic polar coordinates
\end{abstract}
%

\setcounter{equation}{0}
\section{Introduction}
Relations between the geometry of a region~$\Omega$ in~$\Real^n$,
boundary conditions at~$\partial\Omega$, and spectral properties
of the corresponding Laplacian are one of the vintage problems of
mathematical physics. Recent years brought new motivations and
focused attention to aspects of the problem which attracted little
attention earlier.

A strong impetus comes from mesoscopic physics, where new
experimental techniques make it possible to fabricate
semiconductor systems which can be regarded with a reasonable
degree of accuracy as waveguides, resonators, etc., for
effectively free quantum particles. Often potential barriers at
their boundaries can be modeled as a hard wall, in which case it
is natural to identify the system Hamiltonian -- up to a constant
which is usually unimportant -- with the Dirichlet Laplacian,
$-\Delta_D^\Omega$, defined as the Friedrichs extension --
\cf~Section~\ref{Sec.Hamiltonian}. Moreover, the mentioned
solid-state physics advances inspired new insights into the
classical physics, because analogous problems involving Dirichlet
Laplacian arise also in flat electromagnetic waveguides. For more
information about the physical background see ~\cite{DE1,LCM} and
references therein.

On the mathematical side a new interesting effect is the binding
due to the curvature, supposed to be nonzero and asymptotically
vanishing, of an infinitely stretched tubular region in~$\Real^n$,
$n=2,3$. Such ``trapped modes'' may be generated by
other local perturbations of a straight tube as well -- see, \eg,
\cite{BGRS} -- but in the bent-tube case they are of a purely
quantum origin because there are no classical closed trajectories,
apart of a zero measure set of initial conditions in the phase space.

More generally, quantum motion in the vicinity of a manifold with
a potential constraint or Dirichlet condition were studied long
time ago~\cite{JK,daC1,daC2,Tol} in formal attempts to justify
quantization on submanifolds. For a thin neighbourhood one excludes
the transverse part of the Hamiltonian which gives rise to normal
oscillations and the Hamiltonian is replaced by a tangential
operator on the submanifold with the energy appropriately
renormalized. Interest to this problem has been renewed recently
when time evolution around a compact $n$-dimensional manifold 
in~$\Real^{n+m}$ was treated in a rigorous way and compared with 
the corresponding classical dynamics~\cite{FrHe}. The confinement was
realized by a harmonic potential transverse to the manifold and
the thin-neighborhood limit was performed by means of a dilation
procedure followed by averaging in the normal direction. If the
normal bundle is trivial, which is the case, \eg, for manifolds of
codimension one, the resulting tangential Hamiltonian contains two
terms; the first is proportional to the Laplace-Beltrami operator
on the constraint manifold and the second is an effective
potential which depends not only on the intrinsic quantities, but
also on the external curvature of the constraint manifold. Notice
also that if~$\Real^{n+m}$ is replaced by a manifold of the same
dimension, the effective potential depends also on the curvature
of this ambient space~\cite{Mitchell}.

The said potential is important also in the situation when the
width of the ``fat manifold'' is finite and fixed. This was first
noticed for bent planar Dirichlet strips in the paper~\cite{ES1}
which was followed by numerous studies on which the existence
conditions and properties of the geometrically induced discrete
spectrum were further investigated -- see, in particular,
\cite{GJ,DE1,RB}, the first two papers also for 
a generalization to curved tubes in~$\Real^3$. 
On the other hand, much less is known about other possible
generalizations of this problem to higher dimensions starting from
the physically interesting case of curved layers in~$\Real^3$.

This is the question we address in the present paper. While the
strategy will be the same as in the work mentioned above, using
suitable curvilinear coordinates to transform the Laplacian, the
two-dimensional character of the underlying manifold bring new
features. To characterize them briefly, recall that in the
simplest~$(1+1)$-case the effective potential
is~$-\frac{1}{4}\kappa^2$, where~$\kappa$ is the curvature, which
is negative whenever the curvature is nonzero. In case of a layer,
$n=2$ and $m=1$, which we consider here, the (leading term of the)
effective potential is given by~$-\frac{1}{4}(k_1-k_2)^2$ -- see
the derivation of~(\ref{HamiltonianDecoupled}) -- where~$k_1,k_2$
are the principal curvatures of the surface. This expression may
vanish also if the surface is locally spherical, $k_1=k_2$, but
the last relation cannot be valid everywhere at a non-compact
surface unless the latter is a plane, $k_1=k_2=0$. Thus the
effective potential has again an attractive component, which now
combines with a more complicated tangential operator -- the
surface Laplace-Beltrami -- since in distinction to a curve the
surface cannot be fully rectified. This makes the layer case
richer and more interesting.

\setcounter{equation}{0}
\section{Survey of the Paper}
The ultimate objective of this work is to set a list of sufficient conditions
to guarantee the existence of curvature-induced bound states.
We restrict ourselves naturally to non-compact layers only, since
the spectrum of the Dirichlet Laplacian in a bounded region of~$\Real^n$
is always discrete~\cite[Chap.~6]{Davies}.  

The layer configuration space~$\Omega$ itself is properly defined 
in Section~\ref{Sec.Preliminaries} as a tubular neighbourhood of width~$d$
built over a surface~$\Sigma$ embedded in~$\Real^3$
which is diffeomorphic to~$\Real^2$.
To make it more visual, we can understand~$\Omega$ as a part of~$\Real^3$
between a pair of parallel surfaces.
From technical reasons we suppose from the beginning that the surface 
admits at least one \emph{pole} from which we can parametrize 
the surface globally by geodesic polar coordinates.
We stress already here that the existence of a pole in~$\Sigma$ 
is a strong geometric assumption and that there may be 
no poles in general~\cite{GrMe}.
We introduce first quantities describing the layer geometry and 
formulate some basic assumptions. In the subsequent part, 
the Dirichlet Laplacian, $-\Delta_D^\Omega$, is expressed
in terms of the couple~$q=(q^1,q^2)$ of the surface 
(called also longitudinal) coordinates 
together with the normal (transverse) coordinate~$u$.

In Section~\ref{Sec.EssSp}, we estimate the threshold of the essential
spectrum of the Hamiltonian under the assumption~\AssFlat\/ that 
the reference surface is \emph{asymptotically planar} in the sense that 
its Gauss and mean curvatures vanish at large distances. 
We find that this part of spectrum is bounded from below
by~$\kappa_1^2:=\left(\frac{\pi}{d}\right)^2$, which is the lowest
transverse-mode energy.

Section~\ref{Sec.DiscSp} is dedicated to the analysis of the discrete
part of the spectrum. We find here three sufficient conditions and
illustrate them on examples. Since these results leave
open the existence question for thick layers of positive
total Gauss curvature, we present in Section~\ref{Sec.Symmetry} 
an alternative method, which covers the case of asymptotically planar 
layers that are cylindrically symmetric. 
Finally, we conclude in Section~\ref{Sec.Counter} 
by an example of a layer which has no bound states; 
the reference surface here is not asymptotically planar.

To state here the main results of the paper we need to mention
some assumptions which will be discussed in more detail
below:~\AssK\/ and~\AssM\/ means respectively the integrability 
of the Gauss curvature~$K$ and the square of~$\nabla_{\!g} M$, 
where~$M$ is the mean curvature, and~\AssWidth\/
requires the layer half-width to be less than the minimum
normal curvature radius of~$\Sigma$. The integral (total) curvatures
corresponding to~$K$ and~$M$ are defined in~(\ref{Def.Tot}).
\begin{thm}\label{Theorem}
Let~$\Sigma$ be a~$\Smooth^2$-smooth complete simply connected 
non-com\-pact surface with a pole embedded in~$\Real^3$. 
Let the layer~$\Omega$ built over the surface
be not self-intersecting.
If the surface is not a plane but it is asymptotically planar,
then any of the conditions
\begin{itemize}
\item[$\circ$] 
	\AssK\/ and the \emph{total Gauss curvature is non-positive}
\item[$\circ$] 
	$\Sigma$ is $\Smooth^3$-smooth and 
      	the layer is \emph{sufficiently thin}
\item[$\circ$] 
	$\Sigma$ is $\Smooth^3$-smooth, \AssK, \AssM,
      	and the \emph{total mean curvature is infinite}  
\item[$\circ$] 
	\AssK\/ and $\Sigma$ is \emph{cylindrically symmetric}
\end{itemize}
is sufficient for the Laplace operator~$-\Delta_D^\Omega$
to have at least one isolated eigenvalue of finite multiplicity
below~$\inf\sigma_\mathrm{ess}(-\Delta_D^\Omega)$
for all the layer half-widths satisfying~\AssWidth.
\end{thm}

While this theorem covers various wide classes of layers, the list
is not exhaustive. For instance, it remains to be clarified whether
one can include also thick layers without cylindrical symmetry 
built over surfaces with strictly positive total Gauss curvature which, 
however, do not satisfy the assumption~\AssM\/. 
Another open question is whether one can replace~\AssK\/ 
by an assumption including the existence of the total Gauss curvature only,
defined in the principal value sense. 
Finally, it is desirable to find existence results also for layers
over more general surfaces which do not possess poles
or are not diffeomorphic to~$\Real^2$. 

Properties of the obtained curvature-induced bound states
will be discussed elsewhere. Let us just mention that 
in analogy to bent strips~\cite{DE1} one can perform 
the Birman-Schwinger analysis for slightly curved planar layers 
(weak-coupling regime) which yields the first term in the asymptotic 
expansion for the gap between the eigenvalue and the threshold 
of the essential spectrum. We also remark that 
the weak coupling analysis of bent ``fat'' manifolds is similar
to that of a local one-sided deformation
of a straight strip~\cite{BGRS} or planar layer~\cite{BEGK}.

We use the standard component notation of the tensor analysis,
the range of indices being~$1,2$ for Greek and~$1,2,3$ for Latin. 
The indices are associated with the above mentioned coordinates by 
$(1,2,3)\leftrightarrow (q^1,q^2,u)\equiv(s,\vartheta,u)$.
The partial derivatives are denoted by~\emph{commas}, however,
we use also the~\emph{dot} notation for the derivatives w.r.t.~$s$.

\setcounter{equation}{0}
\section{Preliminaries}\label{Sec.Preliminaries}
Let~$\Sigma$ be a~$\Smooth^2$-smooth surface in~$\Real^3$
which has at least one \emph{pole}, \ie, a point~$o\in\Sigma$ 
such that the exponential mapping,
$\exp_o:\tangent_o\Sigma\to\Sigma$, is a diffeomorphism.
The existence of a pole in~$\Sigma$ is a nontrivial assumption
which has important topological consequences. 
In particular, $\Sigma$ is necessarily diffeomorphic to~$\Real^2$ and
as such it is simply connected and non-compact.
Using the \emph{geodesic polar coordinates} 
we can parametrize the surface
(with exception of the pole~$o$) by a unique patch
$p:\Sigma_0\to\Real^3$, where $\Sigma_0:=(0,\infty)\times S^1$.
The tangent vectors $p_{,\mu}:=\partial p/\partial q^\mu$
are linearly independent and their cross-product defines a unit 
normal field~$n$ on~$\Sigma$.
 
Put~$\Omega_0:=\Sigma_0\times (-a,a)$.
We define a \emph{layer}~$\Omega:=\layer(\Omega_0)$ 
of width \mbox{$d=2 a>0$} over the surface~$\Sigma$ by virtue of the mapping
$\layer:\Omega_0\to\Real^3$ which acts as
(\cf~\cite[Prob.~12 of Chap.~3]{Spivak3})
\begin{equation}\label{layer}
  \layer(q,u):=p(q)+u n(q).
\end{equation}
%

\subsection{The Surface Geometry}
The induced surface metric in the geodesic polar coordinates
has the diagonal form, $(g_{\mu\nu})=\diag(1,r^2)$, where
$r^2\equiv g:=\det(g_{\mu\nu})$ is the square of the Jacobian
of the exponential mapping which satisfies the classical Jacobi equation 
\begin{equation}\label{Jacobi}
  \ddot{r}(s,\vartheta)+K(s,\vartheta)\,r(s,\vartheta)=0
  \qquad
  \textrm{with}
  \qquad
  r(0,\vartheta)=0,\ \dot{r}(0,\vartheta)=1.
\end{equation}
The \emph{Gauss curvature}~$K$, together with 
the \emph{mean curvature}~$M$, can be determined
via the Weingarten tensor~$h_\mu^{\ \nu}$ -- \cf~\cite[Prop.~3.5.5]{Kli}.

By means of the invariant surface element,
$d\Sigma:=g^\frac{1}{2}d^2q$,
we may introduce some global quantities characterizing~$\Sigma$,
namely the \emph{total Gauss curvature}~$\TotK$ 
and the \emph{total mean curvature}~$\TotM$
which are defined, respectively, by the integrals
\begin{equation}\label{Def.Tot}
  \TotK:=\int_\Sigma K d\Sigma
  \qquad\textrm{and}\qquad
  \TotM^2:=\int_\Sigma M^2 d\Sigma.
\end{equation}
The latter always exists (it may be $+\infty$), while the former
is well defined provided
\begin{description}
\item\framebox{{\AssK}\quad $K\in\si(\Sigma_0,d\Sigma)$} 
\end{description}
If this condition is not satisfied, one can understand the above integral 
as the principal-value defined through the area restricted
by the geodesic circle~$p(s,\cdot)$ of radius~$s\to\infty$. 
Assuming~$\TotK$ to be finite, an integration of~(\ref{Jacobi}) 
yields the following useful estimate
\begin{equation}\label{r<s}
  \exists C>0\ \forall s\in(0,\infty):
  \quad
  \int_0^{2\pi} r(s,\vartheta)\,d\vartheta \leq C s.
\end{equation}
The norm and the inner product in the Hilbert space 
$L^2(\Sigma_0,d\Sigma)$ will be indicated by the subscript~``$g$''.

\subsection{The Layer Geometry}
It is clear from the definition~(\ref{layer}) that the metric tensor
of the layer (as a manifold with boundary in~$\Real^3$)
has the block form  
\begin{equation}\label{matrixG}
  (G_{ij})=
\begin{pmatrix}
  (G_{\mu\nu}) & 0 \\
  0            & 1 \\
\end{pmatrix}
  \quad\textrm{with}\quad 
  G_{\nu\mu}=
  (\delta_\nu^\sigma-u h_\nu^{\ \sigma})
  (\delta_\sigma^\rho-u h_\sigma^{\ \rho})
  g_{\rho\mu}.
\end{equation}
This formula is well suited for calculation of the determinant,
$G:=\det(G_{ij})$, because the eigenvalues of the matrix 
of the Weingarten map are the \emph{principal curvatures}~$k_1,k_2$, 
and $K=k_1 k_2$, $M=\frac{1}{2}(k_1+k_2)$. Hence
\begin{equation}\label{detG}
  G=g\left[(1-uk_1)(1-uk_2)\right]^2
  =g(1-2Mu+Ku^2)^2.
\end{equation}
In particular, this expression defines through
$d\Omega:=G^\frac{1}{2}d^2q\,du$ the volume element of~$\Omega$.  

Henceforth, we shall assume 
\begin{description}
\item\framebox{{\Ass}\quad $\Omega$ is not self-intersecting} 
	\quad\ie, $\layer$ is injective.  
\end{description}
We have to require also that~$\layer$ is a diffeomorphism.  
In view of the regularity assumptions imposed on~$\Sigma$ and
the inverse function theorem, it is equivalent to assuming
that~$1-2Mu+Ku^2$ does not vanish on~$\Omega_0$, which can be guaranteed
by imposing a restriction on the layer thickness:
\begin{description}
\item\framebox{{\AssWidth}\quad $a<\rho_m:=
	\left(\max\left\{\|k_1\|_\infty,\|k_2\|_\infty\right\}\right)^{-1}$}
\end{description}
The number~$\rho_m$ is naturally interpreted 
as the \emph{minimal normal curvature radius} of~$\Sigma$
(for planar surfaces one can put~$\rho_m:=\infty$).
It follows from~(\ref{matrixG}) that~$C_-\leq 1-2Mu+Ku^2\leq C_+$
holds with~$C_\pm:=\left(1\pm a\rho_m^{-1}\right)^2$.
The lower bound explains why we assume~\AssWidth\/ 
(together with~\Ass) to get the global diffeomorphism. 
On the other hand, the supremum norms in the definition of~$\rho_m$
are necessarily finite since a meaningful layer
must have a non-zero width.  
Another consequence of the considerations is
that under the assumption~\AssWidth\/, 
$G_{\mu\nu}$ can be immediately estimated by the surface metric,
\begin{equation}\label{C+-}
  C_- g_{\mu\nu}\leq G_{\mu\nu}\leq C_+ g_{\mu\nu}
  \qquad\textrm{with}\qquad
  0<C_-\leq 1\leq C_+<4.
\end{equation}
\begin{rem}
We stress the following which will be supposed 
through all the paper but will not be always referred to hereafter:
\begin{description}
\item[$\circ$] 
We consider surfaces which can be parametrized 
by means of the geodesic polar coordinates.
This requires the existence of at least one pole.
\item[$\circ$] 
Since~$\Sigma$ is assumed to be of class~$\Smooth^2$,
the surface curvatures~$K,M$ are~$\Smooth^0$ and as such
bounded locally.
\item[$\circ$] 
Moreover, since we assume layers with non-zero widths,
the principal curvatures have to be bounded uniformly 
on all~$\Sigma_0$ due to~\AssWidth. By virtue of the relation
between~$k_1,k_2$ and~$K,M$, the same is true for the latter.  
\end{description}
\end{rem}
%

\subsection{The Hamiltonian}\label{Sec.Hamiltonian}
After these geometric preliminaries let us define the Hamiltonian 
of our model. We consider a nonrelativistic spinless particle
confined to~$\Omega$ which is free within it and suppose that
the boundary of the layer is a hard wall, \ie, the wavefunctions
satisfy the Dirichlet boundary condition there.
For the sake of simplicity we set Planck's constant $\hbar=1$
and the mass of the particle $m=\frac{1}{2}$. 
Then the Hamiltonian can be identified with the Dirichlet Laplacian
$-\Delta_D^\Omega$ on~$\sii(\Omega)$,
which is defined for an open set $\Omega\subset\Real^3$ as
the Friedrichs extension of the free Laplacian with the domain
defined initially on~$\Comp(\Omega)$ 
-- \cf~\cite[Sec.~XIII.15]{RS4} or~\cite[Chap.~6]{Davies}.
The domain of the closure of the corresponding quadratic form
is the Sobolev space~$\sobi(\Omega)$.

A natural way to investigate this operator is
to pass to the coordinates~$(q,u)$ in which it acquires
the Laplace-Beltrami form ($G_{ij}G^{jk}:=\delta_i^k$)
\begin{equation}\label{Hamiltonian}
  H:=-G^{-\frac{1}{2}}\partial_i G^{\frac{1}{2}}G^{ij}\partial_j
  \qquad
  \textrm{on}
  \qquad
  \sii(\Omega_0,G^\frac{1}{2} d^2q\,du).
\end{equation}
This coordinate change is nothing else than the unitary transformation
$$
  U:\sii(\Omega)\to\sii(\Omega_0,d\Omega):
  \{\psi\mapsto U\psi:=\psi\circ\!\layer\}
$$
which relates the two operators by $H=U(-\Delta_D^\Omega)U^{-1}$.
If~$\Sigma$ is not~$\Smooth^3$-smooth, the operator~$H$ has to
be understood in the form sense
\begin{equation}\label{HamiltonianForm}
  Q[\psi]:=\|H^\frac{1}{2}\psi\|_G^2
  =(\psi_{,i},G^{ij}\psi_{,j})_G, \quad   
  \Dom Q=\sobi(\Omega_0,d\Omega).
\end{equation}
Here the subscript~``$G$'' indicates the norm and the inner product 
in the Hilbert space of~(\ref{Hamiltonian}).
Employing the block form~(\ref{matrixG}) of~$G_{ij}$,
we can split~$H$ into a sum of two parts, $H=H_1+H_2$, given by
\begin{eqnarray}
H_1
& := &-G^{-\frac{1}{2}}\partial_\mu G^{\frac{1}{2}}
  G^{\mu\nu}\partial_\nu
  =-\partial_\mu G^{\mu\nu}\partial_\nu
  -2 F_{,\mu} G^{\mu\nu} \partial_\nu \label{H1} \\
H_2
& := & -G^{-\frac{1}{2}}\partial_3 G^{\frac{1}{2}}\partial_3
  =-\partial_3^2-2\,\frac{Ku-M}{1-2Mu+Ku^2}\,\partial_3\,, \label{H2}
\end{eqnarray}
where we have introduced~$F:=\ln G^\frac{1}{4}$
and expressed~$F_{,3}$ explicitly for~$H_2$ .
 
At the same time, it is useful to have an alternative form 
of the Hamiltonian which has the factor $1-2Mu+Ku^2$ removed from 
the weight~$G^\frac{1}{2}$ of the inner product.
It is obtained by another unitary transformation,
$$
  \hat{U}:\sii(\Omega_0,d\Omega)\to\sii(\Omega_0,d\Sigma\,du):
  \{\psi\mapsto\hat{U}\psi:=(1-2Mu+Ku^2)^\frac{1}{2}\psi\},
$$
which leads to the unitarily equivalent operator
$\hat{H}:=\hat{U}H\hat{U}^{-1}$. This operator makes sense 
if we impose a stronger regularity assumption on~$\Sigma$,
namely that the latter is piecewise~$\Smooth^4$-smooth
(or~$\Smooth^3$ if~$\hat{H}$ is considered in the form sense).  
The operator~$\hat{H}$ can be rewritten by means
of an effective potential~$V$ using $J:=\frac{1}{2}\ln(1-2Mu+Ku^2)$
as follows
$$
  \hat{H}=-g^{-\frac{1}{2}}\partial_i g^{\frac{1}{2}}G^{ij}\partial_j
  +V,
  \qquad
  V
  =g^{-\frac{1}{2}}(g^\frac{1}{2}G^{ij} J_{,j})_{,i}
  +J_{,i}G^{ij}J_{,j}
$$
and again, employing the particular form of~$G_{ij}$, 
the operator~$\hat{H}$ can be split into a sum, 
$\hat{H}_1+\hat{H}_2$. The first operator is defined
by the part of~$\hat{H}$ where one sums over the Greek indices and
$$
  \hat{H}_2=-\partial_3^2+V_2,
  \qquad
  V_2=\frac{K-M^2}{(1-2Mu+Ku^2)^2}.  
$$

To motivate the considerations of the following sections 
let us look at this transformed operator from a heuristic point if view. 
While the operator~$\hat{H}_1+V_2$ depends on all the three coordinates,
in thin layers ($a\ll\rho_m$) its leading term depends
up to an error~$\mathcal{O}(a\rho_m^{-1})$ on the longitudinal
coordinates~$q$ only. 
One can estimate the former in the form sense by means of~(\ref{C+-}) 
and use the fact that $C_\pm=1+\mathcal{O}(a\rho_m^{-1})$.
The transverse coordinate~$u$ is
isolated in~$\hat{H}_2-V_2=-\partial_3^2$, so up to higher-order terms
in~$a$ the Hamiltonian decouples into a sum of the operators
\begin{equation}\label{HamiltonianDecoupled}
  H_q:=-g^{-\frac{1}{2}}\partial_\mu g^{\frac{1}{2}}
  g^{\mu\nu}\partial_\nu+K-M^2
  \qquad
  \textrm{and}
  \qquad
  H_u:=-\partial_3^2\,,
\end{equation}
the first one being the Laplace-Beltrami operator of~$\Sigma$,
except for the additional potential~$K-M^2$ which can be rewritten 
by means of the principal curvatures as~$-\frac{1}{4}(k_1-k_2)^2$.
This is the attractive interaction mentioned in the introduction.
Let us remark that similar Laplace-Beltrami
operators penalized by a quadratic function of the curvature lead
on compact surfaces to interesting isoperimetric
problems~\cite{Ha, HaLo, EHL, Freitas}.

In what follows we shall use 
the family of eigenfunctions~$\{\chi_n\}_{n=1}^\infty$ 
of the transverse operator~$(-\partial_3^2)_D$ which is given by
$$
  \chi_n:=
\begin{cases}
  \sqrt{\frac{2}{d}}\cos\kappa_n u & \textrm{if $n$ is odd}, \\
  \sqrt{\frac{2}{d}}\sin\kappa_n u & \textrm{if $n$ is even}. 
\end{cases}
$$
Here~$\kappa_n^2:=(\kappa_1 n)^2$ with~$\kappa_1:=\pi/d$
are the corresponding eigenvalues.

\setcounter{equation}{0}
\section{Essential Spectrum}\label{Sec.EssSp}
The essential spectrum of a planar layer ($K,M\equiv 0$)
is clearly $[\kappa_1^2,\infty)$. 
By a bracketing argument~\cite[Sec.~3.1]{DEK} and using an appropriate
Weyl sequence, it is easy to see that the same remains true if~$\Omega$ 
is obtained by a compactly supported deformation of a planar layer.
In this section we will prove the inclusion
$\sigma_\mathrm{ess}(-\Delta_D^\Omega)\subseteq[\kappa_1^2,\infty)$
under the assumption that the surface~$\Sigma$ is 
\emph{asymptotically planar} in the sense 
\begin{description}
\item\framebox{{\AssFlat}\quad $K,M\to0$ \qquad as \ $s\to\infty$}
\end{description}
\begin{thm}\label{Thm.EssSp}
Suppose~\emph{\Ass}, \emph{\AssWidth} and assume that the surface 
is asymptotically planar~\emph{\AssFlat}. Then
$$
  \inf\sigma_\mathrm{ess}(-\Delta_D^\Omega)\geq\kappa_1^2.
$$
\end{thm}
\begin{PROOF}
We divide the layer~$\Omega$ into an exterior and interior part
by putting $\Omega_\mathrm{ext}:=\layer(\Omega_{0,s_0})$ 
and $\Omega_\mathrm{int}:=\Omega\setminus\overline{\Omega}_\mathrm{ext}$, 
respectively, where $\Omega_{0,s_0}:=\Sigma_{0,s_0}\times(-a,a)$,
$\Sigma_{0,s_0}:=(s_0,\infty)\times S^1$ for some~$s_0>0$. 
Imposing the Neumann boundary condition at the common boundary
of the two parts, $s=s_0$, we arrive at the decoupled Hamiltonian
$H^N=H_\mathrm{int}^N\oplus H_\mathrm{ext}^N$. More precisely,
it is obtained as the operator associated with the quadratic 
form~$Q^N$ acting as~(\ref{HamiltonianForm}), however with the domain
$\Dom Q^N:=\Dom Q_\mathrm{int}^N\oplus\Dom Q_\mathrm{ext}^N$ where
$$
  \Dom Q_\omega^N
  :=\{\psi\in\Sobi(\Omega_\omega,d\Omega)\,|\ \psi(\cdot,\pm a)=0\},
  \qquad
  \omega\in\{\mathrm{int},\mathrm{ext}\}.
$$
Since $H\geq H^N$ and the spectrum of~$H_\mathrm{int}^N$ 
is purely discrete~\cite[Chap.~7]{Davies},
the minimax principle gives the estimate
$\inf\sigma_\mathrm{ess}(H)\geq\inf\sigma_\mathrm{ess}(H_\mathrm{ext}^N)
\geq\inf\sigma(H_\mathrm{ext}^N)$. Hence it is sufficient to find
a lower bound on~$H_\mathrm{ext}^N$.
However, by virtue of~(\ref{HamiltonianForm}) and~(\ref{matrixG}),
we have for all~$\psi\in\Dom Q_\mathrm{ext}^N$:
\begin{eqnarray*}
  Q_\mathrm{ext}^N[\psi]
&\geq& \|\psi_{,3}\|_{G,\mathrm{ext}}^2
  \geq \inf_{\Omega_{0,s_0}}\{1-2Mu+Ku^2\}\,
  \|\psi_{,3}\|_{\sii(\Omega_0,d\Sigma du),\mathrm{ext}}^2 \\
&\geq& \Big(1-\sup_{\Sigma_{0,s_0}}\{2a|M|+a^2|K|\}\Big)
  \,\kappa_1^2\,\|\psi\|_{\sii(\Omega_0,d\Sigma du),\mathrm{ext}}^2 \\
&\geq& \frac{1-\sup_{\Sigma_{0,s_0}}\{2a|M|+a^2|K|\}}
  {1+\sup_{\Sigma_{0,s_0}}\{2a|M|+a^2|K|\}}
  \,\kappa_1^2\,\|\psi\|_{G,\mathrm{ext}}^2 \\
&=:& \left(1+\epsilon(s_0)\right)\,\kappa_1^2\,\|\psi\|_{G,\mathrm{ext}}^2,
\end{eqnarray*}
where~$\epsilon$ denotes a function which goes to zero as~$s_0\to\infty$ 
due to~\AssFlat. The subscript~``$\mathrm{ext}$'' indicates the restriction 
of the norm to the exterior part. 
In the second line we have used~$(-\partial_3^2)_D\geq\kappa_1^2$. 
The claim then easily follows by the fact that~$s_0$ can be chosen 
arbitrarily large.
\end{PROOF}
\begin{rem}
This threshold estimate is sufficient for the subsequent
investigation of the discrete spectrum which is our goal in this paper. 
In order to show that all energies above $\kappa_1^2$ belong
to the spectrum, one has to construct an appropriate Weyl sequence
to check the opposite conclusion 
$\sigma_\mathrm{ess}(-\Delta_D^\Omega)\supseteq[\kappa_1^2,\infty)$.
This can be done under an assumption stronger than~\AssFlat\/ 
which involves derivatives of the Weingarten tensor as well.
\end{rem}

\setcounter{equation}{0}
\section{Discrete Spectrum}\label{Sec.DiscSp}
The aim of this section is to prove three different conditions
sufficient for the Hamiltonian to have a non-empty spectrum 
below~$\kappa_1^2$. Since we have shown that the essential spectrum 
does not start below this value 
for the layers built over asymptotically planar surfaces,
the conditions yields immediately the existence of curvature-induced
bound states. All the proofs here are based on the variational idea
of finding a trial function~$\Psi$ from the form domain of~$H$ such that
$$
  \Tilde{Q}[\Psi]:=Q[\Psi]-\kappa_1^2\,\|\Psi\|_G^2<0.
$$
It is convenient to split~$Q$ into two parts, $Q=Q_1+Q_2$,
which are associated with~$H_1$ and~$H_2$ of~(\ref{H1})
and~(\ref{H2}), respectively. 

A powerful method in these situation is to construct a trial function 
by deforming the transverse-threshold resonance wavefunction 
separately in the central and tail regions. The idea goes back 
to Goldstone and Jaffe~\cite{GJ}, 
see also~\cite[Thm.~2.1]{DE1},~\cite{RB} and \cite[Sec.~3.2]{DEK}. 
\begin{thm}\label{Thm.GJ}
Assume~\emph{\Ass}, \emph{\AssWidth}, \emph{\AssK},
and suppose that~$\Sigma$ is not planar. If the surface has a non-positive 
total Gauss curvature, \ie, $\TotK\leq 0$, then
$$
  \inf\sigma(-\Delta_D^\Omega)<\kappa_1^2.
$$
\end{thm}
\begin{PROOF}
We begin the construction of~$\Psi$ by considering a radially 
symmetric function $\psi(s,\vartheta,u):=\varphi(s)\chi_1(u)$ 
where~$\varphi$ is arbitrary for a moment. 
Employing the explicit form~(\ref{H2}) of~$H_2$ we get immediately
\begin{equation}\label{GJ-Q2}
  Q_2[\psi]-\kappa_1^2 \|\psi\|_G^2=(\varphi,K\varphi)_g,
\end{equation}
while the ``longitudinal kinetic part''~$Q_1(\psi)$ can be estimated 
by virtue of~(\ref{C+-}) and~(\ref{r<s}) as
\begin{equation}\label{GJ-Q1}
  Q_1[\psi]\leq C_1 \int_0^\infty |\dot{\varphi}(s)|^2 s\,ds.
\end{equation}
The r.h.s. of this inequality depends on the surface geometry 
through the constant~$C_1:=(C_+/C_-)^2 C$ only.
To make this integral arbitrarily small we replace~$\varphi$ by the family 
$\{\varphi_\sigma:\sigma\in(0,1]\}$ of elements which are equal to~$1$
on a compact set, $s\leq s_0$, for some~$s_0>0$, and outside 
they are given by scaled Macdonald functions~\cite[Sec.~9.6]{AS}: 
$$
  \varphi_\sigma(s)
  :=\min\left\{1,\frac{K_0(\sigma s)}{K_0(\sigma s_0)}\right\}.
$$
Since $K_0$ is strictly decreasing, the corresponding 
$\psi_\sigma:=\varphi_\sigma\chi_1$ will not be smooth 
at~$s=s_0$ but it remains continuous, hence it is an admissible 
trial function as an element of~$\Dom Q$. 
Using the properties of the Macdonald function~\cite[Sec.~9.6]{AS} 
and~\cite[5.54]{GR}, it is now easy to verify that
for~$\sigma s_0$ small enough
\begin{equation}\label{DerPhi}
  \exists C_2>0: \qquad
  \int_0^\infty|\dot{\varphi}_\sigma(s)|^2 s\,ds
  <\frac{C_2}{|\ln\sigma s_0|}
\end{equation}
and therefore~$Q_1[\psi_\sigma]\to 0+$ as~$\sigma\to 0+$.
On the other hand, since we assume~\AssK\/ and $|\varphi_\sigma|\leq 1$ 
together with $\varphi_\sigma\to 1-$ pointwise as~$\sigma\to 0+$, 
we get by the dominated convergence theorem that~(\ref{GJ-Q2})
(after the replacement~$\psi\mapsto\psi_\sigma$) 
converges to~$\TotK$. Thus, by choosing~$\sigma$ small enough,
$\tilde{Q}[\psi_\sigma]$ can be made strictly negative if
the total Gauss curvature is strictly negative too.

In order to deal with the case~$\TotK=0$, in analogy to~\cite{GJ} 
we construct the trial function by a small deformation of~$\psi_\sigma$
in the central region.
We set $\Psi_{\sigma,\varepsilon}:=\psi_\sigma+\varepsilon\Theta$
where~$\Theta(q,u):=j(q)u\chi_1(u)$ 
with~$j\in\Comp((0,s_0)\times S^1)$.
Since~$\Theta$ is evidently a function from~$\Dom Q$ as well, we can write
\begin{equation}\label{MixedForm}
  \tilde{Q}[\Psi_{\sigma,\varepsilon}]
  =\tilde{Q}[\psi_\sigma]+2\varepsilon\tilde{Q}(\Theta,\psi_\sigma)
  +\varepsilon^2\tilde{Q}[\Theta]. 
\end{equation}
An explicit calculation where one employs the fact that the scaling acts out 
of the support of the localization function~$j$ yields: 
$\tilde{Q}(\Theta,\psi_\sigma)=-(j,M)_g$,
which can be made non-zero by choosing~$j$ supported on a compact
where~$M$ does not change sign. 
Let us stress that it is independent of~$\sigma$, 
because $\varphi_\sigma=1$ on~$\supp j$; 
the same is true for~$\tilde{Q}[\Theta]$. 
Now such a compact surely exists because
it is supposed that~$\Sigma$ is not a plane and we can take
the parameter~$s_0$ arbitrarily large.
If we choose now the sign of~$\varepsilon$ in such a way that 
the second term on the r.h.s. of~(\ref{MixedForm}) 
is negative, then also the sum with the last term will be negative 
for sufficiently small~$\varepsilon$, and we can choose~$\sigma$ so small 
that~$\tilde{Q}(\Psi_{\sigma,\varepsilon})<0$ 
because~$\tilde{Q}(\psi_\sigma)\to\TotK=0$ as~$\sigma\to 0+$ here.
\end{PROOF}
\begin{rems}
(a) The special choice of the Macdonald function~$K_0$ for 
the mollifier~$\varphi$ is not indispensable. 
In analogy to~\cite{GJ} or~\cite[Thm.~2.1]{DE1} we need
a family of suitable functions scaled exterior to~$(0,s_0)$
in such a way that the integral~(\ref{GJ-Q1}) tends to zero
as~$\sigma\to 0+$. However, since this integral contains 
the extra factor~$s$ (the relic of integration in a higher dimension)
we have to be more careful about the decay properties. 
We have adopted for this purpose the mollifier employed in~\cite{EV,BCEZ}, 
which is the most natural in a sense, because it employs the Green
function kernel of the free 2-dimensional Laplacian at zero energy. 
Nevertheless, we would have succeeded
equally if we had chosen for the scaled tail, \eg,
a compactly supported function similar to that
of the proof of Theorem~\ref{Thm.Symmetry}. 

(b) In the case~$\TotK=0$ we have not used the deformation
proposed in~\cite{DE1}: 
$\tilde{\Theta}:=\tilde{\jmath}^2(H-\kappa_1^2)\psi_\sigma$
with $\tilde{\jmath}\in\Comp((0,s_0)\times S^1\times(-a,a))$,
because it requires an extra condition on the surface regularity.
The analogous condition in the strip case has been forgotten
in~\cite[Thm.~2.1]{DE1}. Moreover, the localization function~$j$
used here is simpler since it is independent of~$u$.
\end{rems}

A class of layers to which the above theorem applies is
represented by those built over \emph{Cartan-Hadamard surfaces},
\ie, geodesically complete simply connected non-compact surfaces 
with non-positive Gauss curvature. 
In view of the Cartan-Hadamard theorem~\cite[Thm.~6.6.4]{Kli} 
each point is a pole and we can therefore construct infinitely
many geodesic polar coordinate systems. Excluding the trivial planar
case, the total Gauss curvature is always strictly negative and so
all these layers possess at least one bound state provided
they are asymptotically planar, $\TotK$ is finite, 
and the assumptions~\Ass,~\AssWidth\/ are satisfied. 
\begin{Ex}[Hyperbolic Paraboloid]
The simple quadric given in~$\Real^3$ by the equation
$z=x^2-y^2$ is an asymptotically planar surface 
with $\TotK=-2\pi$. 
\end{Ex}
\begin{Ex}[Monkey Saddle]
Take $z=x^3-3xy^2$. 
One can again check that~\AssFlat\/ holds true 
and the total Gauss curvature now equals~$-4\pi$.  
\end{Ex}

A family of layers of the limit case~$\TotK=0$ 
was investigated in~\cite{DEK}. We consider there
compactly supported deformations of a planar layer
for which the zero value of~$\TotK$ follows at once
by the Gauss-Bonnet theorem.
If such a deformed plane contains at least one pole,
all the spectral results are trivial consequences 
of the present Theorems~\ref{Thm.EssSp} and~\ref{Thm.GJ}.
On the other hand, the results of~\cite{DEK} are more general 
in the sense that due to the compact support assumption
the technique works without the requirement 
on the existence of a pole.    
\begin{Ex}[Compactly Perturbed Plane without Poles]\label{Ex.Pole}
Suppose that a plane with a circular hole is connected
via a cylindrical tube perpendicular to it with a pierced sphere.
Both interfaces can be made as smooth as needed.
If the tube is sufficiently long there is only one pole~$o$ 
provided the surface has a cylindrical symmetry
w.r.t. the axis of the tube; it coincides with the intersection
of the axis with the sphere.
If we break now the symmetry by taking an ellipsoid
instead of the sphere, we destroy the injectivity of the exponential  
mapping~$\exp_o$ without creating new poles. 
\end{Ex}
\medskip
The Goldstone-Jaffe trick of choosing the ground state of 
the transverse operator as the generalized annulator of the shifted 
energy form~$\tilde{Q}$ has proven its usefulness as a robust argument 
for demonstrating the existence of bound states. 
However, in the present context it reaches its limits because 
the above proof does not work for layers built over surfaces
with positive total curvature, for instance:  
\begin{Ex}[Elliptic Paraboloid]
The surfaces $z=(x/x_0)^2+(y/y_0)^2$ with $x_0,y_0>0$ are
asymptotically planar but~$\TotK=2\pi>0$. They always contain
two poles given by its umbilics which coincide 
if it is a paraboloid of revolution. 
\end{Ex}
\noindent
On the other hand, due to the heuristic argument based 
on~(\ref{HamiltonianDecoupled}) one expects existence
of bound states in any non-planar layer thin enough. 
This is indeed true. This fact together with another
sufficient condition are established in the next theorem.
\begin{thm}\label{Thm.Thin}
Assume~\emph{\Ass},~\emph{\AssWidth}, and suppose that 
$\Sigma$ is \mbox{$\Smooth^3$-smooth}, non-planar and obeys in addition
\emph{
\begin{description}
\item\framebox{{\AssM}\quad 
  $\nabla_{\!g}M\in\sii(\Sigma_0,d\Sigma)$}
\end{description}
}\noindent
Then~$\inf\sigma(-\Delta_D^\Omega)<\kappa_1^2$ if one of the following
two conditions is satisfied:
\begin{itemize}
\item[\emph{(a)}] 
	the layer is sufficiently thin, \ie, $d$ is small enough, 
\item[\emph{(b)}]
	\AssK\/ and the total mean curvature is infinite, \ie,
	$\TotM=\infty$.
\end{itemize}
\end{thm}
\noindent
For brevity we have introduced here the non-component 
notation~$\nabla_{\!g}$ for the covariant derivative on~$\Sigma$.
\begin{PROOF}
We use
$\Psi_\sigma(s,\vartheta,u):=\left(1+M(s,\vartheta)u\right)\psi_\sigma(s,u)$,
where~$\psi_\sigma=\varphi_\sigma\chi_1$ is the trial function 
defined in the first part of the proof of Theorem~\ref{Thm.GJ}. 
Under the stated regularity assumption, $\Psi_\sigma$ 
is an admissible trial function, \ie, it belongs to~$\Dom Q$. 
Using~(\ref{C+-}) together with Minkovski's inequality and~(\ref{H2}), 
we get 
\begin{align*}
  Q_1[\Psi_\sigma]
&\leq 2(C_+/C_-)^2
  \left(\left(1+a\|M\|_\infty\right)^2\|\dot{\varphi}_\sigma\|_g^2 
  +a^2\|\varphi_\sigma\nabla_{\!g}M\|_g^2\right) \\
  Q_2[\Psi_\sigma]-\kappa_1^2 \|\Psi_\sigma\|_G^2
&= \left(\varphi_\sigma,(K-M^2)\varphi_\sigma\right)_g
  +\frac{\pi^2-6}{12\kappa_1^2}
  \left(\varphi_\sigma,K M^2\varphi_\sigma\right)_g.
\end{align*}
We start by checking the second sufficient condition. We recall that
due to~\AssWidth\/, $K$ and~$M$ are uniformly bounded.
Thus, thanks to~\AssM\/ and the hypotheses assumed in~(b), 
it follows that~$\tilde{Q}[\Psi_\sigma]\to-\infty$ as~$\sigma\to 0+$.

We pass now to the first sufficient condition.
Since~$K-M^2$ is negative -- \cf~(\ref{HamiltonianDecoupled}) -- 
continuous and the surface is supposed to be non-planar, 
the first term at the r.h.s. of the second line is strictly negative,
say~$-c^2$, for sufficiently large value of~$s_0$ (the radius of the disc
where~$\psi_\sigma=\chi_1$). On the other hand, 
$\|\dot{\varphi}_\sigma\|_g$ is estimated by~(\ref{DerPhi}), so we can
choose~$\sigma$ so small that it is less than~$c^2/3$. 
Now we choose the layer half-width~$a$ so small that the sum of the
remaining terms of the estimated~$\tilde{Q}[\Psi_\sigma]$ is less 
than~$c^2/3$ as well. For this we recall that~$\kappa_1^{-2}$
is proportional to~$a^2$.
Hence~$\tilde{Q}[\Psi_\sigma]\leq -c^2/3<0$ for~$\sigma,d$ small enough. 
\end{PROOF}
\begin{rem}
In order to obtain the first sufficient condition,
one can replace \AssM\/ by an assumption 
on the boundedness of~$\nabla_{\!g}M$. Moreover,
if we had used the compactly supported function~$\varphi_n$ 
from the proof of Theorem~\ref{Thm.Symmetry} below
instead of~$\varphi_\sigma$,
it would have been sufficient to assume that~$\nabla_{\!g}M$
was bounded locally only, which is exactly the situation
when~$\Sigma$ is of class~$\Smooth^3$. This is why~\AssM\/
is not included in the thin layer case of Theorem~\ref{Theorem}.
\end{rem}

We believe that the hypothesis~\AssM\/ is technical 
-- \cf~Example~\ref{Ex.M}.
Even with it, however, the class of layers possessing 
bound states without any restriction on the layer thickness
other than~\AssWidth\/ is extended significantly. 
For instance, it is an easy exercise to verify that all 
the conditions of Theorem~\ref{Thm.Thin}~(b) are fulfilled for the elliptic 
paraboloids and many other surfaces with a positive total Gauss curvature. 
Removing this technical condition is still an open question 
except for layers endowed with the cylindrical symmetry which 
we shall discuss below. 

\setcounter{equation}{0}
\section{Cylindrically Symmetric Layers}\label{Sec.Symmetry}
Consider now layers which are invariant w.r.t. rotations around 
a fixed axis in~$\Real^3$. We may thus suppose that~$\Sigma$ is a surface 
of revolution parametrized by~$p:\Sigma_0\to\Real^3$,
$$
  p(s,\vartheta):=
  \left(r(s)\cos\vartheta,r(s)\sin\vartheta,z(s)\right),
  \qquad\textrm{where}\quad
  r,z\in\Smooth^2\left((0,\infty)\right),\ r>0.
$$
It will be the geodesic polar coordinate chart if we impose 
the following condition on the \emph{canonical} parametrization, 
\begin{equation}\label{canonical}
  \dot{r}^2+\dot{z}^2=1;
  \qquad
  \textrm{then also}
  \qquad
  \dot{r}\ddot{r}+\dot{z}\ddot{z}=0.
\end{equation}
An explicit calculation yields the diagonal form
of the Weingarten tensor, $(h_\mu^{\ \nu})=\diag(k_s,k_\vartheta)$,
with the principal curvatures
$k_s=\dot{r}\ddot{z}-\ddot{r}\dot{z}$ and \mbox{$k_\vartheta=\dot{z}r^{-1}$}.
In fact, it is sufficient to know the function~$s\mapsto k_s(s)$ only, 
since~$r,z$ can be constructed from the relations
\begin{equation}\label{Surface.Construction}
\begin{aligned}
  r(s)&=\int_0^s\cos b(\xi)\,d\xi \\
  z(s)&=\int_0^s\sin b(\xi)\,d\xi
\end{aligned}
  \qquad\quad\textrm{with}\qquad
  b(s):=\int_0^{s} k_s(\xi)\,d\xi.
\end{equation}

Recall that by~Theorem~\ref{Thm.GJ} the spectrum bottom of any layer
is strictly less than the first transverse eigenvalue provided~$\TotK\leq 0$. 
However, only the case~$\TotK=0$ is relevant to the present situation
of surfaces of revolution, because by the Gauss-Bonnet theorem
(see~also~(\ref{Jacobi})) 
\begin{equation}\label{Jacobi.Symmetry}
  \TotK+2\pi \dot{r}(\infty)=2\pi,
  \qquad\textrm{where}\quad
  \dot{r}(\infty):=\lim_{s\to\infty}r(s), 
\end{equation}
and~$\dot{r}(\infty)>1$ is not allowed because of~(\ref{canonical}). 
Notice, on the other hand, that~$\dot{r}(\infty)$
always exists since the existence of the total Gauss curvature is supposed.
Moreover, the positivity of~$r$ requires~$\TotK\leq 2\pi$.

The goal of this section is to show that in the present special case 
of symmetric layers $\inf\sigma(-\Delta_D^\Omega)<\kappa_1^2$  
holds true also for all admissible strictly positive values of~$\TotK$,
irrespective of the layer thickness.
Our argument requires to exclude here the extreme case $\TotK=0$
for which the result is already known, without any symmetry assumption.
Hereafter we will therefore assume that~$0\leq\dot{r}(\infty)<1$.
It follows that there exist~$0<\delta'<\frac{1}{2}$ 
and~$s_0>0$ such that for all~$s\geq s_0$ 
one has~$-\delta'\leq\dot{r}(s)\leq 1-\delta'$. 
Using now the explicit dependence of~$k_\vartheta$ on~$r,\dot{z}$ 
and~(\ref{canonical}), we obtain the essential ingredients of our strategy:
\begin{lemma}\label{Lemma.Ingredients}
Assume~$\TotK>0$.
There exist~$\delta>0$ and~$s_0>0$ such that
$$
  \forall s\geq s_0:\quad
  \frac{\delta}{r(s)}\leq |k_\vartheta(s)| \leq \frac{1}{r(s)}
  \quad\mbox{and}\quad
  k_\vartheta(s)\ \mbox{does not change sign}.
$$
\end{lemma}
\noindent
In particular, employing~(\ref{r<s}), it follows that~$k_\vartheta$ is
not integrable in~$\si(\posReal)$. On the other hand, 
the meridian curvature~$k_s$ is integrable under the assumption~\AssK, 
which is seen by the regularity properties imposed on~$p$ 
and the following estimate
$$
  \infty>\int_0^\infty |K(s)|\,r(s)\,ds
  \geq\int_{s_0}^\infty |k_s(s) k_\vartheta(s)|\,r(s)\,ds 
  \geq\delta\int_{s_0}^\infty |k_s(s)|\,ds. 
$$

This is the essence of what we are going to use
in our method. Even if~$M$ may decay at infinity it is not
negligible in the integral sense there. However, 
$K$ is supposed to be integrable and it will enable 
us to eliminate the unpleasant contribution of the corresponding total
curvature -- \cf~(\ref{GJ-Q2}) -- by going to large distances 
by means of a family of trial functions supported there. 
\begin{thm}\label{Thm.Symmetry}
Assume~\emph{\Ass}, \emph{\AssWidth}, \emph{\AssK},
and suppose that~$\Sigma$ is a surface of revolution. Then
$
  \inf\sigma(-\Delta_D^\Omega)<\kappa_1^2.
$
\end{thm}
\begin{PROOF} 
Since the result for~$\TotK=0$ is included in Theorem~\ref{Thm.GJ},
we suppose~$\TotK>0$ in the following.
We use~$\Psi_{n,\varepsilon}(s,u)
:=(\varphi_n(s)+\varepsilon \phi_n(s)u)\chi_1(u)$, where 
$\varepsilon$ will be specified later and~$\varphi_n,\phi_n$ are
functions ``localized at infinity'' as~$n\to\infty$. They are
defined in the following way: Consider three sequences
\mbox{$b_1,b_2,b_3:\Nat\to\Nat$} such that $0<b_1<b_2<b_3$
and~$b_1(n)\to\infty$ as~$n\to\infty$. We set    
$$
  \varphi_n(s):=\frac{\ln(s/b_i)}{\ln(b_j/b_i)},
  \quad
  (i,j)\in\{(1,2),(3,2)\},
  \qquad\textrm{and}\qquad
  \phi_n(s):=\frac{\varphi_n(s)}{s}
$$ 
if $\min\{b_i,b_j\}< s \leq \max\{b_i,b_j\}$,
and assume that~$\varphi_n,\phi_n$ are zero elsewhere.  
Defined in this way the functions are not smooth at the matching points,
however, $\Psi_{n,\varepsilon}$ still belongs to~$\Dom Q$
because they are continuous and of a compact support for
each \mbox{$n\in\Nat$}. Next we note that they
are positive and uniformly bounded 
(the maximum of $\phi_n$ is even decreasing as~$n\to\infty$).

Using~(\ref{C+-}) and~(\ref{r<s}) we can estimate
the longitudinal kinetic parts of~$\tilde{Q}$
-- \cf\/ also~({\ref{GJ-Q1}}) -- by one-dimensional integrals
$$
  Q_1[\varphi_n\chi_1]\leq C_1 
  \int_0^\infty\dot{\varphi}_n(s)^2 s\,ds,
  \qquad
  Q_1[\phi_nu\chi_1]\leq \frac{d^2}{2} C_1 
  \int_0^\infty\dot{\phi}_n(s)^2 s\,ds,
$$
and an explicit calculation yields that both
converge to zero as~$n\to\infty$ if we demand, in addition,
that~$b_2/b_1$ and~$b_3/b_2$ tend to infinity as~$n\to\infty$. 
The same is true for the mixed term~$Q_1(\varphi_n\chi_1,\phi_nu\chi_1)$
by the Schwarz inequality.
On the other hand, an explicit integration w.r.t.~$u$
for the rest of~$\tilde{Q}$ yields
\begin{multline*}
  Q_2[\Psi_{n,\varepsilon}]-\kappa_1^2 \|\Psi_{n,\varepsilon}\|_G \\
  =(\varphi_n,K\varphi_n)_g
  -2\varepsilon(\varphi_n,M\phi_n)_g
  +\varepsilon^2\left[\|\phi_n\|_g^2
  +\frac{\pi^2-6}{3\kappa_1^2}\,(\phi_n,K\phi_n)_g\right].
\end{multline*}
For large~$n$ the contribution of the Gauss curvature will
be negligible because of~\AssK\/ 
and the facts that~$\varphi_n$ and~$\phi_n$ are uniformly bounded
and the infimum of their support tends to infinity as~$n\to\infty$. 
Summing up the results, we arrive at
\begin{equation}\label{OUR-Result}
  \lim_{n\to\infty} \tilde{Q}[\Psi_{n,\varepsilon}]=
  \lim_{n\to\infty} \left[\varepsilon^2\|\phi_n\|_g^2
  -2\varepsilon (\varphi_n,M\phi_n)_g\right] 
\end{equation}
if the limit on the r.h.s. exists.

We put~$\varepsilon\equiv\varepsilon_n:=(\varphi_n,M\phi_n)_g^{-1}$
which will be seen in a moment as a reasonable choice
because the integral tends to infinity as~$n\to\infty$
for particular choices of~$b_j$; $\varepsilon_n$ is thus well-defined 
for~$n$ large enough. Then the problem turns to comparing
the number~$-2$ to the limit
$$
  \lim_{n\to\infty}\frac{(\phi_n,\phi_n)_g}
  {(\varphi_n,M\phi_n)_g^2}.
$$

In the special case of cylindrically symmetric surfaces
when one has the information about the explicit behaviour of~$M$
at infinity, it is an easy matter. Indeed, since~$k_s$
is integrable in~$\si(\posReal)$ and $\phi_n$ is chosen in a way 
to eliminate the weight~$r$ with help of~(\ref{r<s}),
the meridian curvature does not contribute in the denominator, while
in view of~Lemma~\ref{Lemma.Ingredients}, $k_\vartheta r$ 
can be replaced by a constant value near infinity. 
Using in addition~(\ref{r<s}) in the numerator, 
one is therefore seeking the zero limit of 
$$
  \frac{\int_0^\infty\phi_n(s)^2 s\,ds}
  {\left(\int_0^\infty\varphi_n(s)\phi_n(s) ds\right)^2}
  =\frac{1}{\int_0^\infty\phi_n(s)^2 s\,ds}
  =\frac{3}{\ln(b_3/b_1)}.
$$
One can choose, for instance, $\forall n\geq 2$:
$b_1(n):=n$, $b_2(n):=n^2$, $b_3(n):=n^3$, 
which fulfill also the other properties earlier required 
about these sequences.
We conclude by~$\tilde{Q}[\Psi_{n,\varepsilon}]\to -2$ 
as~$n\to\infty$ so we can find a finite~$n_0$ 
for which the form will be negative.
\end{PROOF}
\begin{rem}
Notice that~(\ref{OUR-Result}) is a general result. We have
not supposed anything of the surface symmetry 
when deriving this relation.
\end{rem}
\begin{Ex}[Hyperboloid of Revolution]
Consider one of the two sheets of the hyperboloid given by
the equation~$x^2+y^2-(z/z_0)^2=1$. 
It is an asymptotically planar surface of revolution 
and via the parameter~$z_0>0$ we can get arbitrary value
of the total Gauss curvature between~$0$ and~$2\pi$.  
\end{Ex}
\begin{Ex}[Surface with Non Square Integrable~$\nabla_{\!g}M$]\label{Ex.M}
Let us con\-struct an asymptotically planar surface
of revolution which satisfies~\AssK\/ but contradicts~\AssM.
We define~$k_s(s):=s^{-2}\sin s^2$ and use~(\ref{Surface.Construction})
to get the functions~$r,z$ and in this way the map~$p$. 
One can easily check that there is a~$c$ such that~$r(s)\geq c s$
for all~$s\in\posReal$. Therefore $k_\vartheta=\dot{z}r^{-1}\to 0$ 
as~$s\to\infty$ because~$|\dot{z}|=|\sin b(s)|\leq 1$; 
the same limit holds, of course, for~$k_s$.
Since~$K,M$ are expressed by means of the principal curvatures,
it follows that the surface is asymptotically planar~\AssFlat.
At the same time, $|K|r=|k_s\dot{z}|\leq |k_s|$ is integrable
in~$\si(\posReal)$ which gives~\AssK. On the other hand,
while it is true that~$\dot{k}_\vartheta=k_s r^{-1}\cos b-r^{-2}\sin b\cos b$ 
belongs to~$\sii(\posReal,r(s)ds)$, 
the same does not hold for~$\dot{k}_s$ by its definition.
Hence, $\nabla_{\!g}M=(\dot{M},0)$ does not fulfil~\AssM.
We note that an explicit calculation 
together with~(\ref{Jacobi.Symmetry}) 
yields~$\TotK=2\pi\left(1-\cos\sqrt{\frac{\pi}{2}}\right)
\thickapprox 1.38\,\pi$ in this example.
\end{Ex}
\begin{rem}(\emph{Partial Wave Decomposition}).
An alternative approach is to decompose~$-\Delta_D^\Omega$ 
with respect to angular momentum subspaces
to investigate the spectral
properties of layers endowed with the cylindrical symmetry. 
The obtained series of partial-wave Hamiltonians
have similar form as the pure strip Hamiltonian -- \cf~\cite{ES1,DE1} --
except for an additional centrifugal term and different operator domain
for the lowest wave. This, however, makes the spectral analysis
of layers more complicated than a direct use of the non-decomposed 
Hamiltonian~$H$. At the same time, it gives an insight 
into the choice of the trial function in the proof 
of Theorem~\ref{Thm.Symmetry} which has to be supported in the region
where the influence of the centrifugal term is negligible. 
\end{rem}

\setcounter{equation}{0}
\section{A Layer without Bound States}
\label{Sec.Counter}
Consider a semi-cylinder of radius~$R$ closed by a hemisphere; 
the total Gauss curvature is~$2\pi$. Since the mean curvature
of the cylindrical part is constant,~$M=(2R)^{-1}>0$,
such a surface is not asymptotically planar.  
We shall demonstrate that the Hamiltonian~$H:=-\Delta_D^\Omega$
of the corresponding layer~$\Omega$ built over this surface 
does not possess bound states for any~$a<R$.

Imposing the Neumann or Dirichlet boundary condition 
on the segment of connection of the hemispherical 
and cylindrical layer, we get the bounds
$
  H_\mathit{sph}^N\oplus H_\mathit{cyl}^N 
  \leq H \leq
  H_\mathit{sph}^D\oplus H_\mathit{cyl}^D
$.
The spectrum of the hemispherical-segment Hamiltonians is purely discrete.
By the minimax principle only the cylindrical part
of the estimating operators contributes to the essential spectrum,
while a possible eigenvalue of~$H$ below the essential spectrum is squeezed
between the corresponding eigenvalues 
of~$H_\mathit{sph}^N$ and~$H_\mathit{sph}^D$.
In particular, for our purpose it is sufficient to show that
$\inf\sigma(H_\mathit{sph}^N)>\inf\sigma_\mathrm{ess}(H_\mathit{cyl}^D)$.
The spectral analysis of these operators becomes trivial
if they are expressed in the spherical or cylindrical coordinates,
respectively. 

Due to the mirror symmetry, the ground state energy
of~$H_\mathit{sph}^N$ is the same as the lowest eigenvalue 
of the entire spherical layer which is~$\kappa_1^2 $.
On the other hand, $\sigma(H_\mathit{cyl}^j)
=\sigma_\mathrm{ess}(H_\mathit{cyl}^j)=[\epsilon_1,\infty)$
for both the conditions $j\in\{N,D\}$, 
where the threshold~$\epsilon_1$ is given by the first eigenvalue
of the radial operator~$-\partial_r^2-(4 r^2)^{-1}$ on~$\sii(\posReal)$.
Since the latter is less than $-\partial_r^2-(4(R+a)^2)^{-1}$,
the Rayleigh principle yields~$\epsilon_1<\kappa_1^2$.
It is now easy to conclude that the spectrum of the unified layer satisfies
\begin{equation}\label{Sp.Counter}
  \sigma(H)=\sigma_\mathrm{ess}(H)=[\epsilon_1,\infty).
\end{equation}

\begin{rem}
The above example shows that without the condition~\AssFlat\/,
or at least without~$M\to 0$ at the infinity, 
one cannot guarantee the existence of bound states. 
Notice that the reference surface is not~$\Smooth^2$-smooth 
in this counter-example and thus it does not belong to the class 
of manifolds considered from the beginning. 
Nevertheless, one can construct a sequence of domains which
converges in an appropriate sense 
to the hemispherical layer and, at the same time,
they can be connected to the cylindrical part 
in a sufficiently smooth way. It follows then from~\cite[Thm.~1.5]{RT}
that the spectral result~(\ref{Sp.Counter}) remains preserved
for the domains sufficiently close to the limiting layer.
\end{rem}

\subsection*{Acknowledgments}
The authors would like to thank for private communications
to Mark S.~Ashbaugh, and to Wolfgang T.~Meyer who
suggested Example~\ref{Ex.Pole}.
The work has been done during the visits of~P.~E. and~D.~K.
to Centre de Physique Th\'eorique, Marseille-Luminy,
and~P.~D. to the Nuclear Physics Institute, AS~CR; 
the authors express their gratitude to the hosts.
The work has been partially supported by the Grant AS A~1048101
and the CAS-CNRS Exchange Agreement~7919.
%
%
\providecommand{\bysame}{\leavevmode\hbox to3em{\hrulefill}\thinspace}

\end{document}